\documentclass[preprint,superscriptaddress]{revtex4-1} 
\usepackage{graphicx} 
\usepackage{amsmath, cases} 
\usepackage[version=3]{mhchem} 
\usepackage{longtable} 
\usepackage{filecontents} 
\usepackage{cleveref} 
\usepackage{multirow} 
\usepackage{subfigure} 

\usepackage{booktabs,siunitx,array,threeparttable} 
\usepackage{mathtools} 
\usepackage{url} 
\usepackage{verbatim} 
 
\usepackage{epstopdf} 
\usepackage{esint} 
\usepackage{cancel} 
\usepackage{amssymb} 
\usepackage{upgreek}

\usepackage{color}
\usepackage[normalem]{ulem}

\begin{document} 

\title{Spin Symmetry in Thermally-Assisted-Occupation Density Functional Theory} 

\author{Yu-Yang Wang} 
\affiliation{Department of Physics, National Taiwan University, Taipei 10617, Taiwan} 

\author{Jeng-Da Chai} 
\email[Author to whom correspondence should be addressed. Electronic mail: ]{jdchai@phys.ntu.edu.tw} 
\affiliation{Department of Physics, National Taiwan University, Taipei 10617, Taiwan} 
\affiliation{Center for Theoretical Physics and Center for Quantum Science and Engineering, National Taiwan University, Taipei 10617, Taiwan} 
\affiliation{Physics Division, National Center for Theoretical Sciences, Taipei 10617, Taiwan}

\date{\today} 

\begin{abstract} 

For electronic systems with multi-reference (MR) character, Kohn-Sham density functional theory (KS-DFT) with the conventional exchange-correlation (xc) energy functionals can lead to incorrect spin densities and related 
properties. For example, for H$_{2}$ dissociation, the spin-restricted and spin-unrestricted solutions obtained with the same xc energy functional in KS-DFT can be distinctly different, yielding the unphysical spin-symmetry 
breaking effects in the spin-unrestricted solutions. Recently, thermally-assisted-occupation density functional theory (TAO-DFT) has been shown to resolve the aforementioned spin-symmetry breaking, when the fictitious 
temperature is properly chosen. In this work, a response theory based on TAO-DFT is developed to demonstrate that TAO-DFT with a sufficiently large fictitious temperature can always resolve the unphysical 
spin-symmetry breaking in MR systems. To further support this, TAO-DFT calculations with various fictitious temperatures are performed for the dissociation of H$_{2}$, N$_{2}$, He$_{2}$, and Ne$_{2}$ as well as the 
twisted ethylene. 

\end{abstract}

\maketitle

\section{Introduction} 

In condensed matter physics and quantum chemistry, the challenge of accurately predicting the properties of electronic systems at affordable computational cost has led to the development of numerous electronic structure 
methods \cite{Jensen_2007}. Among all the competitors, Kohn-Sham density function theory (KS-DFT) \cite{Hohenberg_Kohn_1964,Kohn_Sham_1965} has become the most popular electronic structure method. By 
introducing a noninteracting auxiliary system at absolute zero, KS-DFT successfully bypasses the challenge of expressing the exact form of the kinetic energy functional, resulting in the successful predictions of the 
ground-state properties of electronic systems at relatively low computational cost \cite{Parr_Yang_1989,Kummel_Kronik_2008}. 

Nevertheless, KS-DFT with the conventional exchange-correlation (xc) energy functionals can yield erroneous results when dealing with electronic systems with multi-reference (MR) character (also called MR systems or 
strongly correlated electron systems) \cite{SciYang}. Recognizing this limitation, thermally-assisted-occupation density functional theory (TAO-DFT) \cite{Chai_2012} has recently emerged as an alternative. Instead of utilizing 
the ground-state density of a noninteracting auxiliary system at absolute zero as in KS-DFT, TAO-DFT adopts the thermal equilibrium density of a noninteracting auxiliary system at a non-zero temperature (the so-called 
fictitious temperature) $\theta$ in its formulation. This allows the ground-state density of an electronic system to be represented with the orbitals and their occupations in TAO-DFT. The introduction of fractional orbital 
occupations enables TAO-DFT to effectively correct some of the unphysical results of KS-DFT in MR systems \cite{Chai_2012,Chai_2014,Chai_2017}. 

One of the well-known challenges in KS-DFT is the dissociation of H$_{2}$ in the spin-unrestricted calculations, where the results deviate from the spin-restricted results due to the emergence of unphysical spin-symmetry 
breaking. In the exact theory, due to spin symmetry, the spatial distribution of the two spin densities should be identical in the electronic ground state \cite{Helgaker_Jorgensen_Olsen_2000}. When the H$_{2}$ bond is 
stretched beyond the Coulson-Fischer (CF) point \cite{CFPoint}, the solutions obtained with spin-unrestricted calculations differ from those obtained with spin-restricted calculations. In other words, the up-spin and down-spin densities become unequal and the degeneracy of KS orbitals breaks down beyond the CF point.
This unphysical spin-symmetry breaking feature stems from the MR character of H$_{2}$ at the dissociation limit. Being incapable of dealing with MR character, KS-DFT with 
the conventional xc energy functionals fails to obtain the correct spin-unrestricted predictions, although the KS-DFT solutions are more stable than the Hartree-Fock (HF) solutions, as discovered by the stability analysis of Bauernschmitt and Ahlrichs \cite{Bauernschmitt_Ahlrichs_1996}. 
Ensemble DFT is one way to solve this symmetry breaking problem \cite{Nagy_1998}. In TAO-DFT, previous
numerical investigations have shown that the spin symmetry of H$_{2}$ dissociation can be restored in TAO-DFT with a sufficiently large fictitious 
temperature $\theta$ \cite{Chai_2012}. What remains unknown is whether this restoration 
of spin symmetry is a system-independent behavior in TAO-DFT. This underscores the need for a theory capable of characterizing the impact of fictitious temperature on the spin symmetry in TAO-DFT. 

In this work, our theory is introduced in Sec.~\ref{sec:theory}, which is divided into four parts. In Sec.~\ref{sec:iteration}, we develop a response theory within the TAO-DFT framework. 
This response theory provides a more straightforward way to observe the fictitious-temperature dependence of the spin-symmetry breaking. 
Based on the response theory, in Sec.~\ref{sec:criterion}, we establish a criterion to determine whether the spin symmetry is restored. 
The explicit forms of the response theory are derived in Sec.~\ref{sec:kernel}--\ref{sec:basis}.
In Sec.~\ref{sec:limit}, we study the asymptotic behavior in the high-fictitious-temperature limit. 
Numerical investigations of our theory are provided in Sec.~\ref{sec:calculation}. 
Our conclusions are provided in Sec.~\ref{sec:conclusions}.

\section{\label{sec:theory} Theory}

\subsection{\label{sec:iteration} Iteration of KS Equations}

In TAO-DFT, the spin densities of the electronic ground state are determined by the following self-consistent equations \cite{Chai_2012}: 
\begin{align} 
    v_{\text{S}}^{\sigma}(\mathbf{r}) 
    =&\ v(\mathbf{r}) + \frac{\delta\left( {E_{\text{H}} + E_{\text{xc}\theta} } \right)}{\delta\rho^{\sigma}(\mathbf{r})}, 
    \label{eq:scf_v} \\
    \left[ {- \frac{\nabla^{2}}{2} + v_{\text{S}}^{\sigma}(\mathbf{r})} \right]\psi_{i}^{\sigma}(\mathbf{r})
    =&\ \varepsilon_{i}^{\sigma}\psi_{i}^{\sigma}(\mathbf{r}), 
    \label{eq:scf_psi} \\ 
    f_{i}^{\sigma} 
    =&\ \frac{1}{1 + {\exp\left[ {\left( {\varepsilon_{i}^{\sigma} - \mu^{\sigma}} \right)/\theta} \right]}}, 
    \label{eq:scf_f} \\ 
    N^{\sigma} 
    =&\ {\sum_{i}f_{i}^{\sigma}}, 
    \label{eq:scf_N} \\ 
    \rho^{\sigma}(\mathbf{r}) 
    =&\ {\sum_{i}{f_{i}^{\sigma}\left| {\psi_{i}^{\sigma}(\mathbf{r})} \right|^{2}}}. 
    \label{eq:scf_rho}
\end{align}
Here, $\sigma$ = $\alpha$ or $\beta$ denotes the up-spin or down-spin, $v_{\text{S}}^{\sigma}$ is the $\sigma$-spin effective one-electron potential, $v$ is the external potential, $E_{\text{H}}$ is the Hartree energy, and $E_{\text{xc}\theta}$ is the exchange-correlation-$\theta$ (xc$\theta$) energy (conventional $E_\text{xc}$ plus $E_\theta$) \cite{Chai_2012}, $\mu^{\sigma}$ is the $\sigma$-spin chemical potential used to adjust the Fermi-Dirac occupation numbers $\{f_{i}^{\sigma}\}$ to satisfy the sum rule given by Eq.~(\ref{eq:scf_N}).
Each cycle of self-consistent equations (Eqs.~(\ref{eq:scf_v})--(\ref{eq:scf_rho})) maps the input spin densities $(\rho^{\alpha},\rho^{\beta})$ to the output spin densities $(\rho_{\text{out}}^{\alpha},\rho_{\text{out}}^{\beta})$.
We can obtain the damped output spin densities $(\rho_{\text{damped}}^{\alpha},\rho_{\text{damped}}^{\beta})$ by mixing the input and output spin densities,
\begin{align}
    \rho_{\text{damped}}^{\sigma}(\mathbf{r}) = t\rho_{\text{out}}^{\sigma}(\mathbf{r})+(1-t)\rho^{\sigma}(\mathbf{r}),
    \label{eq:scf_damping}
\end{align}
where $0<t\leq 1$. This also serves as the input spin densities in the next cycle.
Repeat iterating the self-consistent equations (Eqs.~(\ref{eq:scf_v})--(\ref{eq:scf_rho})) where Eq.~(\ref{eq:scf_damping}) with small enough $t$ is used, then we obtained a sequence of spin densities whose convergence is guaranteed \cite{Wagner_2013}.
Although several options of self-consistent field (SCF) algorithms can be adopted in actual calculations, it is sufficient for us to consider the damping given by Eq.~(\ref{eq:scf_damping}) only, since the criterion of spin-symmetry (Eq.~(\ref{eq:criterion})) derived in the follows does not depend on the SCF algorithm we use.

A pair of converged spin densities is a fixed point of the mapping given by Eq.~(\ref{eq:scf_damping}) since the input spin densities $(\rho^\alpha,\rho^\beta)$ equals the damped output spin densities $(\rho_{\text{damped}}^\alpha,\rho_{\text{damped}}^\beta)$.
For a system possessing a singlet ground state, the converged spin-restricted density is a fixed point of the mapping, since an input with $\rho^{\alpha} = \rho^{\beta}$ cannot result in an output with $\rho^{\alpha}_{\text{damped}} \neq \rho^{\beta}_{\text{damped}}$ by the symmetry argument. 
In fact, the input with equal spin up-spin and down-spin densities gives the same KS-potential (Eq.~(\ref{eq:scf_v})), which is responsible for the same $\{\psi_{i}^{\sigma}\}$ and $\{\varepsilon_{i}^{\sigma}\}$ for up-spin and down-spin (Eq.~(\ref{eq:scf_psi})). By Eqs.~(\ref{eq:scf_f})--(\ref{eq:scf_N}), the degenerate KS orbitals should be equally filled, so the equality $\rho^{\alpha}_{\text{out}} = \rho^{\beta}_{\text{out}}$ as well as $\rho^{\alpha}_{\text{damped}} = \rho^{\beta}_{\text{damped}}$ should be maintained.
To detect the stability of the fixed point, we add a perturbation $\{\delta\rho^{\alpha},\delta\rho^{\beta}\}$ to the converged spin-restricted density to break the spin symmetry, and observe the response in the output spin densities.
It can be obtained by the chain rule and Eqs.~(\ref{eq:scf_v})--(\ref{eq:scf_damping}):
\begin{align}
    \delta\rho_{\text{out}}^{\sigma}(\mathbf{r})
    =&\ \sum_{\sigma'}\int{K^{\sigma\sigma'}(\mathbf{r},\mathbf{r}')\delta\rho^{\sigma'}(\mathbf{r}')\,d\mathbf{r}'},
    \label{eq:iteration_undamped} \\
    \delta\rho_{\text{damped}}^{\sigma}(\mathbf{r})
    =&\ t\sum_{\sigma'}\int{K^{\sigma\sigma'}(\mathbf{r},\mathbf{r}')\delta\rho^{\sigma'}(\mathbf{r}')\,d\mathbf{r}'} + (1-t)\delta\rho^\sigma(\mathbf{r}),
    \label{eq:iteration}
\end{align}
where the kernel of undamped iteration is
\begin{align}
    K^{\sigma\sigma'}(\mathbf{r},\mathbf{r}')
    =&\ \int\left[
    \sum_{i} {\int{\frac{\delta\rho_{\text{out}}^{\sigma}(\mathbf{r})}{\delta\psi_{i}^{\sigma}(\mathbf{r}''')}\frac{\delta\psi_{i}^{\sigma}(\mathbf{r}''')}{\delta v_{\text{S}}^{\sigma}(\mathbf{r}'')}\,d\mathbf{r}'''}} 
    + \sum_{i} {\int{\frac{\delta\rho_{\text{out}}^{\sigma}(\mathbf{r})}{\delta\psi_{i}^{\sigma*}(\mathbf{r}''')}\frac{\delta\psi_{i}^{\sigma*}(\mathbf{r}''')}{\delta v_{\text{S}}^{\sigma}(\mathbf{r}'')}\,d\mathbf{r}'''}} \right.\notag\\ 
    &+ \left.\sum_{ij} \frac{\delta\rho_{\text{out}}^{\sigma}(\mathbf{r})}{\delta f_{i}^{\sigma}} \frac{\delta f_{i}^{\sigma}}{\delta\varepsilon_{j}^{\sigma}} \frac{\delta\varepsilon_{j}^{\sigma}}{\delta v_{\text{S}}^{\sigma}(\mathbf{r}'')}
    \right]
    \frac{\delta v_{\text{S}}^{\sigma}(\mathbf{r}'')}{\delta\rho^{\sigma'}(\mathbf{r}')}\,d\mathbf{r}''.
    \label{eq:kernel_underived} 
\end{align}
We will derive and analyze the expression of Eq.~(\ref{eq:kernel_underived}) to explain the reason why the spin-symmetry breaking can be restored in TAO-DFT.
In Eq.~(\ref{eq:kernel_underived}), $\psi_{i}^{\sigma}$ and $\psi_{i}^{\sigma*}$ are considered as two separated variations, yet we can also treat $\rho_{\text{out}}$ solely as a function of $\psi_{i}^{\sigma}$, with $\delta\rho_{\text{out}}^{\sigma}/\delta\psi_{i}^{\sigma*} = 0$ given as the condition for complex differentiability of $\rho_{\text{out}}$.
Eq.~(\ref{eq:iteration_undamped}) and (\ref{eq:iteration}) can be rewritten by means of the density-perturbation vector $\delta\rho$ and the undamped iteration operator $K$,
\begin{align}
    \delta\rho_{\text{out}} =&\ K\,\delta\rho,
    \label{eq:iteration_operator_undamped} \\
    \delta\rho_{\text{damped}} =&\ [tK+(1-t)]\delta\rho.
    \label{eq:iteration_operator}
\end{align}
Eq.~(\ref{eq:iteration_operator_undamped}) and (\ref{eq:iteration_operator}) can also represent the matrix equation corresponding to Eq.~(\ref{eq:iteration_undamped}) and (\ref{eq:iteration}) on a given basis set respectively.

\subsection{\label{sec:criterion} Criterion of Spin-Symmetry}

In a KS-DFT calculation, the converged spin-restricted density possesses triplet instability if it is unstable with respect to the spin-symmetry breaking perturbations \cite{Bauernschmitt_Ahlrichs_1996,Helgaker_Jorgensen_Olsen_2000,Jensen_2007}.
For a system possessing no triplet instability, the spin-symmetry is preserved in the electronic ground state predicted by the spin-unrestricted calculations if there is no other local minimum of the energy functional.
For such kind of systems, any spin-symmetry breaking density-perturbation vector $\delta\rho$ will eventually vanish under repeated iterations of Eq.~(\ref{eq:iteration_operator}) with a small enough $t=t_{j}$ at each $j$-th step, since iteration of self-consistent equations (Eqs.~(\ref{eq:scf_v})--(\ref{eq:scf_rho})) with sufficient damping converges to the converged spin-restricted density \cite{Wagner_2013,Penz_et_al._2019,Penz_et_al._2020}.
It is expressed in terms of
\begin{align}
    \prod_{j=1}^{\infty}[t_{j}K+(1-t_{j})]\cdot\delta\rho = 0,
    \label{eq:asymp}
\end{align}
where $K$ is determined by Eq.~(\ref{eq:kernel_underived}), $\delta\rho$ is a spin-symmetry breaking perturbation satisfying the sum rule
\begin{align}
    \int{\delta\rho^{\sigma}(\mathbf{r})\,d\mathbf{r}}
    = \delta N^{\sigma}
    = 0.
\end{align}
On the other hand, if Eq.~(\ref{eq:asymp}) hold for every spin-symmetry breaking density-perturbation vector and small enough $\{t_{j}\}_{j=1}^{\infty}$, then the system does not possess triplet instability.

By diagonalization
\begin{align}
    K = P\Lambda P^{-1}, 
\end{align}
where $\Lambda$ is a diagonal matrix, and $P$ is an invertible matrix, we have
\begin{align}
    \prod_{j=1}^{\infty}[t_{j}K+(1-t_{j})]
    =&\ \prod_{j=1}^{\infty}[t_{j}P\Lambda P^{-1}+(1-t_{j})PP^{-1}] \notag\\
    =&\ \prod_{j=1}^{\infty}\{P[t_{j}\Lambda+(1-t_{j})]P^{-1}\} \notag\\
    =&\ P\prod_{j=1}^{\infty}[t_{j}\Lambda+(1-t_{j})]\cdot P^{-1}.
\end{align}
Thus, Eq.~(\ref{eq:asymp}) is equivalent to
\begin{align}
    \prod_{j=1}^{\infty}
    [t_{j}\Lambda+(1-t_{j})]\cdot\delta\rho = 0.
    \label{eq:asymp_diagonal}
\end{align}
Since
\begin{align}
    t_{j}\Lambda+(1-t_{j})
    = \sum_{i}{[t_{j}\lambda_{i}+(1-t_{j})]|e_i\rangle\langle e_i|},
\end{align}
where $\{\lambda_{i}\}$ and $\{|e_i\rangle\}$ are the eigenvalues and eigenvectors of $K$, Eq.~(\ref{eq:asymp_diagonal}) holds if
\begin{align}
    |t\lambda_{i}+(1-t)| < 1
    \label{eq:criterion_underived_1}
\end{align}
for some sufficiently small $t$ and for all $i$ whose $\{|e_i\rangle\}$ contribute to spin-symmetry breaking.
Since that $|z|=1$ is a circle on the complex plane and that $t\lambda_{i}+(1-t)$ is just the linear interpolation between $z=1$ and $z=\lambda_{i}$, Eq.~(\ref{eq:criterion_underived_1}) holds for small enough $t\in(0,1]$ if and only if
\begin{align} 
    \operatorname{Re}{\lambda_{i}} < 1.
    \label{eq:criterion_underived_2}
\end{align}
Define $\lambda$ be the largest real part of eigenvalues that contributes to spin-symmetry breaking.
By Eq.~(\ref{eq:criterion_underived_2}), if
\begin{align}
    \lambda < 1,
    \label{eq:criterion}
\end{align}
then Eq.~(\ref{eq:asymp}) holds, the system does not possess triplet instability, and the spin-symmetry is preserved.
On the contrary, if $\lambda > 1$, then for some $\lambda_{i}$, $\operatorname{Re}{\lambda_{i}} > 1$, $|t\lambda_{i}+(1-t)| > 1$ for any $t$, and Eq.~(\ref{eq:asymp}) cannot be satisfied by any choice of $\{t_{j}\}$; the triplet instability results in the spin-symmetry breaking.
By the criterion Eq.~(\ref{eq:criterion}), the spin-symmetry can be checked by using only the converged spin-restricted data.

\subsection{\label{sec:kernel} Derivation of Kernel of Undamped Iteration}

The terms in Eq.~(\ref{eq:kernel_underived}) are derived with the following steps.
The quantities without any index of spin are the converged spin-restricted values, where up-spin and down-spin quantities are equal.

\begin{enumerate}

\item
For the noninteracting auxiliary system, by the first-order nondegenerate perturbation theory, we obtain \cite{Shankar_2012}
\begin{align}
    \delta\psi_{i}^{\sigma}(\mathbf{r}''')
    =&\ {\sum_{j(\neq i)}{\frac{\left\langle \psi_{j} \middle| {\delta v_{\text{S}}^{\sigma}} \middle| \psi_{i} \right\rangle}{\varepsilon_{i}-\varepsilon_{j}}\psi_{j}(\mathbf{r}''')}} \notag\\
    =&\ {\int{{\sum_{j(\neq i)}\frac{\psi_{j}(\mathbf{r}''')\psi^{*}_{j}(\mathbf{r}'')}{\varepsilon_{i}-\varepsilon_{j}}}\psi_{i}(\mathbf{r}'')\,\delta v_{\text{S}}^{\sigma}(\mathbf{r}'')\,d\mathbf{r}''}}, \\
        \delta\psi_{i}^{\sigma*}(\mathbf{r}''')
    =&\ {\int{{\sum_{j(\neq i)}\frac{\psi^{*}_{j}(\mathbf{r}''')\psi_{j}(\mathbf{r}'')}{\varepsilon_{i}-\varepsilon_{j}}}\psi^{*}_{i}(\mathbf{r}'')\,\delta v_{\text{S}}^{\sigma}(\mathbf{r}'')\,d\mathbf{r}''}}, \\
        \delta\varepsilon_{i}^{\sigma}
    =&\ \left\langle \psi_{i} \middle| {\delta v_{\text{S}}^{\sigma}} \middle| \psi_{i} \right\rangle
    = {\int{| {\psi_{i}(\mathbf{r}'')}|^{2}\delta v_{\text{S}}^{\sigma}(\mathbf{r}'')\,d\mathbf{r}''}},
\end{align}
where $\{\psi_i\}$ and $\{\varepsilon_i\}$ are the converged spin-restricted orbitals and energy eigenvalues. Thus, 
\begin{align}
    \frac{\delta\psi_{i}^{\sigma}(\mathbf{r}''')}{\delta v_{\text{S}}^{\sigma}(\mathbf{r}'')}
    =&\ {\sum_{j(\neq i)}\frac{\psi_{j}(\mathbf{r}''')\psi^{*}_{j}(\mathbf{r}'')}{\varepsilon_{i}-\varepsilon_{j}}}\psi_{i}(\mathbf{r}''),
    \label{eq:derivative_psi_v} \\
    \frac{\delta\psi_{i}^{\sigma*}(\mathbf{r}''')}{\delta v_{\text{S}}^{\sigma}(\mathbf{r}'')}
    =&\ {\sum_{j(\neq i)}\frac{\psi^{*}_{j}(\mathbf{r}''')\psi_{j}(\mathbf{r}'')}{\varepsilon_{i}-\varepsilon_{j}}}\psi^{*}_{i}(\mathbf{r}''),
    \label{eq:derivative_psi_v_star} \\
    \frac{\delta\varepsilon_{i}^{\sigma}}{\delta v_{\text{S}}^{\sigma}(\mathbf{r}'')}
    =&\ |{\psi_{i}(\mathbf{r}'')}|^{2}.
    \label{eq:derivative_epsilon_v}
\end{align}

\item 
By Eq.~(\ref{eq:scf_rho}), 
\begin{align}
\delta\rho_{\text{out}}^{\sigma}(\mathbf{r}) 
= \sum_{i}{\left[
f_{i}\psi^{*}_{i}(\mathbf{r})\,\delta\psi_{i}^{\sigma}(\mathbf{r})
+ f_{i}\psi_{i}(\mathbf{r})\,\delta\psi_{i}^{\sigma*}(\mathbf{r})
+ |{\psi_{i}(\mathbf{r})}|^{2}\,\delta f_{i}^{\sigma} \right]},
\end{align}
where $\{f_i\}$ are the converged spin-restricted occupation numbers. 
Thus,
\begin{align}
    \frac{\delta\rho_{\text{out}}^{\sigma}(\mathbf{r})}{\delta\psi_{i}^{\sigma}(\mathbf{r}''')} 
    =&\ f_{i}\psi_{i}^{*}(\mathbf{r})\,\delta(\mathbf{r}-\mathbf{r}'''),
    \label{eq:derivative_rho_psi} \\
    \frac{\delta\rho_{\text{out}}^{\sigma}(\mathbf{r})}{\delta\psi_{i}^{\sigma*}(\mathbf{r}''')} 
    =&\ f_{i}\psi_{i}(\mathbf{r})\,\delta(\mathbf{r}-\mathbf{r}'''),
    \label{eq:derivative_rho_psi_star} \\
    \frac{\delta\rho_{\text{out}}^{\sigma}(\mathbf{r})}{\delta f_{i}^{\sigma}} 
    =&\ |\psi_i(\mathbf{r})|^{2}.
    \label{eq:derivative_rho_f}
\end{align}

\item 
Since
\begin{align}
\delta f_{i}^{\sigma}
=&\ \delta\left( \frac{1}{1 + {\exp\left[ {\left( {\varepsilon_{i}^{\sigma} - \mu^{\sigma}} \right)/\theta} \right]}} \right) \notag \\ 
=&\ -\frac{\exp\left[ {\left( {\varepsilon_{i} - \mu} \right)/\theta} \right]}{\left\{ {1 + {\exp\left[ {\left( {\varepsilon_{i} - \mu} \right)/\theta} \right]}} \right\}^{2}}\frac{\delta\varepsilon_{i}^{\sigma} - \delta\mu^{\sigma}}{\theta} \notag \\ 
=&\ \frac{1}{\theta}f_{i}\left( {1 - f_{i}} \right)\left( {\delta\mu^{\sigma} - \delta\varepsilon_{i}^{\sigma}} \right), 
\end{align} 
where $\{f_i\}$ and $\{\varepsilon_i\}$ are the converged spin-restricted occupation numbers and energy eigenvalues, by the sum rule 
\begin{align} 
\frac{1}{\theta}{\sum_{i}{f_{i}\left( {1 - f_{i}} \right)\left( {\delta\mu^{\sigma} - \delta\varepsilon_{i}^{\sigma}} \right)}}
= {\sum_{i}{\delta f_{i}^{\sigma}}}
= \delta N^{\sigma}
= 0,
\end{align} 
we obtain 
\begin{align} 
\delta\mu^{\sigma} = \frac{\sum_{i}{f_{i}\left( {1 - f_{i}} \right)\delta\varepsilon_{i}^{\sigma}}}{\sum_{i}{f_{i}\left( {1 - f_{i}} \right)}}. 
\end{align} 
Thus, 
\begin{align} 
\delta f_{i}^{\sigma} 
=&\ \frac{1}{\theta}f_{i}\left( {1 - f_{i}} \right)\left[ {\frac{\sum_{j}{f_{j}\left( {1 - f_{j}} \right)\delta\varepsilon_{j}^{\sigma}}}{\sum_{k}{f_{k}\left( {1 - f_{k}} \right)}} - \delta\varepsilon_{i}^{\sigma}} \right]  \notag \\ 
=&\ {\sum_{j}{\frac{1}{\theta}f_{i}\left( {1 - f_{i}} \right)\left[ {\frac{f_{j}\left( {1 - f_{j}} \right)}{\sum_{k}{f_{k}\left( {1 - f_{k}} \right)}} - \delta_{ij}} \right]\delta\varepsilon_{j}^{\sigma}}}.
\end{align} 
For simplicity, we define
\begin{align} 
    \beta_{ij}
    = \frac{\delta f^\sigma_i}{\delta\varepsilon^\sigma_j}
    = \frac{1}{\theta}f_{i}\left( {1 - f_{i}} \right)\left[ {\frac{f_{j}\left( {1 - f_{j}} \right)}{\sum_{k}{f_{k}\left( {1 - f_{k}} \right)}} - \delta_{ij}} \right]. 
    \label{eq:beta}
\end{align}

\item 
We have 
\begin{align} 
\frac{\delta v_{\text{S}}^{\sigma}(\mathbf{r}'')}{\delta\rho^{\sigma'}(\mathbf{r}')} 
=&\ \frac{\delta}{\delta\rho^{\sigma'}(\mathbf{r}')}\left[ {v(\mathbf{r}'') + \frac{\delta\left( {E_{\text{H}} + E_{\text{xc}\theta} } \right)}{\delta\rho^{\sigma}(\mathbf{r}'')}} \right] \notag \\ 
=&\ \frac{\delta E_{\text{H}}}{\delta\rho^{\sigma}(\mathbf{r}'')\delta\rho^{\sigma'}(\mathbf{r}')} + \frac{\delta E_{\text{xc}\theta}}{\delta\rho^{\sigma}(\mathbf{r}'')\delta\rho^{\sigma'}(\mathbf{r}')} \notag \\ 
=&\ \frac{1}{\left|{\mathbf{r}''-\mathbf{r}'}\right|} + g_{\text{xc}\theta}^{\sigma\sigma'}(\mathbf{r}'',\mathbf{r}'), 
\label{eq:derivative_v_rho} 
\end{align} 
in which
\begin{align}
g_{\text{xc}\theta}^{\sigma\sigma'}(\mathbf{r}'',\mathbf{r}') = \frac{\delta E_{\text{xc}\theta}}{\delta\rho^{\sigma}(\mathbf{r}'')\delta\rho^{\sigma'}(\mathbf{r}')}
\end{align}
with value taken on the converged spin-restricted density. 
Specifically, for the local density approximation (LDA), Eq.~(\ref{eq:derivative_v_rho}) is reduced to the form 
\begin{align} 
\frac{\delta v_{\text{S}}^{\sigma}(\mathbf{r}'')}{\delta\rho^{\sigma'}(\mathbf{r}')}
=&\ \frac{1}{\left|{\mathbf{r}''-\mathbf{r}'}\right|} + g_{\text{xc}\theta}^{\sigma\sigma'}(\mathbf{r}')\,\delta(\mathbf{r}''-\mathbf{r}'). 
\label{eq:derivative_v_rho_LDA} 
\end{align} 
\end{enumerate} 


By Eqs.~(\ref{eq:derivative_psi_v}) and (\ref{eq:derivative_rho_psi}), we obtain
\begin{align}
\sum_{i}{\int{\frac{\delta\rho_{\text{out}}^{\sigma}(\mathbf{r})}{\delta\psi_{i}^{\sigma}(\mathbf{r}''')}\frac{\delta\psi_{i}^{\sigma}(\mathbf{r}''')}{\delta v_{\text{S}}^{\sigma}(\mathbf{r}'')}\,d\mathbf{r}'''}}
=&\ \sum_{i}{\int{f_{i}\psi_{i}^{*}(\mathbf{r})\delta(\mathbf{r}-\mathbf{r}'''){\sum_{j(\neq i)}\frac{\psi_{j}(\mathbf{r}''')\psi_{j}^{*}(\mathbf{r}'')}{\varepsilon_{i}-\varepsilon_{j}}}\psi_{i}(\mathbf{r}'')\,d\mathbf{r}'''}} \notag \\ 
=&\ \sum_{ij(i\neq j)}{\frac{f_{i}}{\varepsilon_{i}-\varepsilon_{j}}\psi_{i}^{*}(\mathbf{r})\psi_{j}(\mathbf{r})\psi_{i}(\mathbf{r}'')\psi_{j}^{*}(\mathbf{r}'')}.
\end{align}
Similarly, by Eqs.~(\ref{eq:derivative_psi_v_star}) and (\ref{eq:derivative_rho_psi_star}), we obtain
\begin{align}
\sum_{i}{\int{\frac{\delta\rho_{\text{out}}^{\sigma}(\mathbf{r})}{\delta\psi_{i}^{\sigma*}(\mathbf{r}''')}\frac{\delta\psi_{i}^{\sigma*}(\mathbf{r}''')}{\delta v_{\text{S}}^{\sigma}(\mathbf{r}'')}\,d\mathbf{r}'''}}
=& \sum_{ij(i\neq j)}{\frac{f_{i}}{\varepsilon_{i}-\varepsilon_{j}}\psi_{i}(\mathbf{r})\psi_{j}^{*}(\mathbf{r})\psi_{i}^{*}(\mathbf{r}'')\psi_{j}(\mathbf{r}'')} \notag\\
=&\ \sum_{ij(i\neq j)}{\frac{f_{j}}{\varepsilon_{j}-\varepsilon_{i}}\psi_{i}^{*}(\mathbf{r})\psi_{j}(\mathbf{r})\psi_{i}(\mathbf{r}'')\psi_{j}^{*}(\mathbf{r}'')}.
\end{align}
Thus,
\begin{align}
    &\sum_{i}{\int{\frac{\delta\rho_{\text{out}}^{\sigma}(\mathbf{r})}{\delta\psi_{i}^{\sigma}(\mathbf{r}''')}\frac{\delta\psi_{i}^{\sigma}(\mathbf{r}''')}{\delta v_{\text{S}}^{\sigma}(\mathbf{r}'')}\,d\mathbf{r}'''}}
    + \sum_{i}{\int{\frac{\delta\rho_{\text{out}}^{\sigma}(\mathbf{r})}{\delta\psi_{i}^{\sigma*}(\mathbf{r}''')}\frac{\delta\psi_{i}^{\sigma*}(\mathbf{r}''')}{\delta v_{\text{S}}^{\sigma}(\mathbf{r}'')}\,d\mathbf{r}'''}} \notag\\
    &= \sum_{ij(i\neq j)}{\frac{f_{i}-f_{j}}{\varepsilon_{i}-\varepsilon_{j}}\psi_{i}^{*}(\mathbf{r})\psi_{j}(\mathbf{r})\psi_{i}(\mathbf{r}'')\psi_{j}^{*}(\mathbf{r}'')} \notag \\ 
    &= \sum_{ij}{\alpha_{ij}\psi_{i}^{*}(\mathbf{r})\psi_{j}(\mathbf{r})\psi_{i}(\mathbf{r}'')\psi_{j}^{*}(\mathbf{r}'')},
    \label{eq:derivative_rho_v}
\end{align}
where 
\begin{align} 
    \alpha_{ij} =
    \begin{cases} 
        \displaystyle\frac{f_{i}-f_{j}}{\varepsilon_{i}-\varepsilon_{j}}, &i\neq j; \\
        0, &i=j.
    \end{cases} 
    \label{eq:alpha}
\end{align}
Note that under the limit $\varepsilon_{i}\to\varepsilon_{j}$ ($i\neq j$), $\alpha_{ij}$ gives a finite number $f'(\varepsilon_{i})$ for $\theta\neq 0$.
This enables us to consider a system with degenerate levels as the limit of a sequence of nondegenerate systems.
By taking this limit, Eq.~(\ref{eq:derivative_rho_v}) can be used for systems with degenerate levels, while the fraction $(f_{i}-f_{j})/(\varepsilon_{i}-\varepsilon_{j})$ in Eq.~(\ref{eq:alpha}) is replaced with the derivative $f'(\varepsilon_{i})$.
By Eqs.~(\ref{eq:kernel_underived}), (\ref{eq:derivative_epsilon_v}), (\ref{eq:derivative_rho_f}), (\ref{eq:derivative_v_rho}), and (\ref{eq:derivative_rho_v}), 
we obtain the kernel of undamped iteration
\begin{align}
    K^{\sigma\sigma'}(\mathbf{r},\mathbf{r}')
    =&\ {\int{{\sum_{ij}\begin{Bmatrix}
            \alpha_{ij}\psi^{*}_{i}(\mathbf{r})\psi_{j}(\mathbf{r})\psi_{i}(\mathbf{r}'')\psi^{*}_{j}(\mathbf{r}'') \\
            + \beta_{ij}| {\psi_{j}(\mathbf{r})} |^{2}| {\psi_{i}(\mathbf{r}'')} |^{2}
        \end{Bmatrix}}\left[\frac{1}{\left|{\mathbf{r}''-\mathbf{r}'}\right|} + g_{\text{xc}\theta}^{\sigma\sigma'}(\mathbf{r}'',\mathbf{r}')\right]\,d\mathbf{r}''}} \notag\\
    =&\ {\sum_{ij}\left[ {\alpha_{ij}\psi^{*}_{i}(\mathbf{r})\psi_{j}(\mathbf{r})\eta_{ij}^{\sigma\sigma'}(\mathbf{r}') + \beta_{ij}|{\psi_{j}(\mathbf{r})}|^{2}\eta_{ii}^{\sigma\sigma'}(\mathbf{r}')} \right]}.
    \label{eq:kernel}
\end{align}
where 
\begin{align} 
\eta_{ij}^{\sigma\sigma'}(\mathbf{r}') = {\int{\left[ {\frac{1}{\left|{\mathbf{r}''-\mathbf{r}'}\right|} + g_{\text{xc}\theta}^{\sigma\sigma'}(\mathbf{r}'',\mathbf{r}')} \right]\psi_{i}(\mathbf{r}'')\psi^{*}_{j}(\mathbf{r}'')\,d\mathbf{r}''}}. 
\end{align}

\subsection{\label{sec:basis} Response Theory with a Basis Set}

When a finite basis set $\{\chi_i(\mathbf{r})\}$ is employed, not all $\delta\rho^\sigma(\mathbf{r})$ are accessible in Eq.~(\ref{eq:iteration}).
The perturbations $\delta\rho^{\sigma}(\mathbf{r})$ are attributed to the perturbation of finite number of coefficients, resulting in an easier derivation without involving the interchange of limits.
By expanding the spin density on the basis set, we obtain 
\begin{align} 
\rho^{\sigma}(\mathbf{r})
=&\ \sum_{i}{f_{i}^{\sigma} \left|{\psi_{i}^{\sigma}(\mathbf{r})}\right|^{2}} 
= \sum_{ij}{
        \left(\sum_{k}{f_{k}^{\sigma}c_{ki}^{\sigma*}c_{kj}^{\sigma}}\right)
        \chi_{i}(\mathbf{r})\chi_{j}(\mathbf{r})
    }. 
\end{align}
It is shown that $\{\chi_i(\mathbf{r})\chi_j(\mathbf{r})\}$'s form the basis of the spin densities and that $\delta\rho^\sigma(\mathbf{r})$ should be their linear combination, 
\begin{align}
\delta\rho^{\sigma}(\mathbf{r}) 
= {\sum_{ij}{\delta\rho_{ij}^{\sigma}\,\chi_{i}(\mathbf{r})\chi_{j}(\mathbf{r})}}. 
\label{eq:delta_rho_expansion} 
\end{align} 
The sum rule requires 
\begin{align} 
\sum_{ij}{S_{ij}\delta\rho_{ij}^{\sigma}}
= \int{\delta\rho^{\sigma}(\mathbf{r})\,d\mathbf{r}}
= \delta N^{\sigma}
= 0,
\label{eq:sum_rule_basis} 
\end{align} 
where 
\begin{align} 
S_{ij} 
= {\int{\chi_{i}(\mathbf{r})\chi_{j}(\mathbf{r})\,d\mathbf{r}}}.
\label{eq:matrix_S} 
\end{align} 
Eqs.~(\ref{eq:delta_rho_expansion}) and (\ref{eq:sum_rule_basis}) constitute the constraints of $\delta\rho^\sigma(\mathbf{r})$ when a basis set is employed. 

We rewrite the iteration formula (Eq.~(\ref{eq:iteration})) on a basis set.
By Eq.~(\ref{eq:kernel}), we have
\begin{align}
K^{\sigma\sigma'}(\mathbf{r},\mathbf{r}')
=&\ \sum_{ij} \left[
        \alpha_{ij}\psi_{i}^{*}(\mathbf{r})\psi_{j}(\mathbf{r})\eta_{ij}^{\sigma\sigma'}(\mathbf{r}')
        + \beta_{ij}|{\psi_{j}(\mathbf{r})}|^{2}\eta_{ii}^{\sigma\sigma'}(\mathbf{r}')
    \right] \notag\\
=&\ \sum_{ijmn} \chi_{i}(\mathbf{r})\chi_{j}(\mathbf{r})\left[
        {\alpha_{mn}c_{mi}^{*}c_{nj}\eta_{mn}^{\sigma\sigma'}(\mathbf{r}')
        + \beta_{mn}c_{ni}^{*}c_{nj}\eta_{mm}^{\sigma\sigma'}(\mathbf{r}')}
    \right] \notag\\
=&\ \sum_{ij} \chi_{i}(\mathbf{r})\chi_{j}(\mathbf{r})K_{ij}^{\sigma\sigma'}(\mathbf{r}'),
\end{align}
where 
\begin{align} 
K_{ij}^{\sigma\sigma'}(\mathbf{r}') = {\sum_{mn}\left[ {\alpha_{mn}c_{mi}^{*}c_{nj}\eta_{mn}^{\sigma\sigma'}(\mathbf{r}') + \beta_{mn}c_{ni}^{*}c_{nj}\eta_{mm}^{\sigma\sigma'}(\mathbf{r}')} \right]}. 
\end{align}
By Eqs.~(\ref{eq:iteration}) and (\ref{eq:delta_rho_expansion}), 
we obtain 
\begin{align}
    \delta\rho_{\text{damped}}^{\sigma}(\mathbf{r})
    =&\ t\sum_{\sigma'}\int{K^{\sigma\sigma'}(\mathbf{r},\mathbf{r}')\delta\rho^{\sigma'}(\mathbf{r}')\,d\mathbf{r}'} + (1-t)\delta\rho^\sigma(\mathbf{r}) \notag\\
    =&\ \sum_{ij}\chi_{i}(\mathbf{r})\chi_{j}(\mathbf{r})\left[\,
            t\sum_{\sigma'}{\int{K_{ij}^{\sigma\sigma'}(\mathbf{r}')\,\delta\rho^{\sigma'}(\mathbf{r}')\,d\mathbf{r}'}}+(1-t)\delta\rho_{ij}^{\sigma}
        \right] \notag\\
    =&\ \sum_{ij} \chi_{i}(\mathbf{r})\chi_{j}(\mathbf{r})\left[\,
            t\sum_{\sigma'kl}{K_{ij,kl}^{\sigma\sigma'}\,\delta\rho_{kl}^{\sigma'}}
            + (1-t)\delta\rho_{ij}^{\sigma}
        \right],
\end{align}
where
\begin{align}
    K_{ij,kl}^{\sigma\sigma'}
    =&\ {\int{K_{ij}^{\sigma\sigma'}(\mathbf{r}')\chi_{k}(\mathbf{r}')\chi_{l}(\mathbf{r}')\,d\mathbf{r}'}} \notag\\
    =&\ {\sum_{mn}{\int{\left[ {\alpha_{mn}c_{mi}^{*}c_{nj}\eta_{mn}^{\sigma\sigma'}(\mathbf{r}') + \beta_{mn}c_{ni}^{*}c_{nj}\eta_{mm}^{\sigma\sigma'}(\mathbf{r}')} \right]\chi_{k}(\mathbf{r}')\chi_{l}(\mathbf{r}')\,d\mathbf{r}'}}} \notag\\
    =&\ {\sum_{mnpq}\left[ \left(\alpha_{mn}c_{mi}^{*}c_{nq}^{*}+\beta_{mn}c_{mq}^{*}c_{ni}^{*}\right)c_{mp}c_{nj}\xi_{pq,kl}^{\sigma\sigma'} \right]},
    \label{eq:kernel_basis} \\
    \xi_{pq,kl}^{\sigma\sigma'}
    =&\ \iint
        \chi_{p}(\mathbf{r})\chi_{q}(\mathbf{r}) \left[
            \frac{1}{\left| \mathbf{r}-\mathbf{r}'\right|}
            + g_{\text{xc}\theta}^{\sigma\sigma'}(\mathbf{r},\mathbf{r}')
        \right] \chi_{k}(\mathbf{r}')\chi_{l}(\mathbf{r}')
        \,d\mathbf{r}\,d\mathbf{r}'.
\label{eq:xi} 
\end{align}
Specifically, for the LDA, Eq.~(\ref{eq:xi}) reduces to the form 
\begin{align} 
\xi_{pq,kl}^{\sigma\sigma'}
= \int
        {\chi_{p}(\mathbf{r})\chi_{q}(\mathbf{r}) 
        g_{\text{xc}\theta}^{\sigma\sigma'}(\mathbf{r})
        \chi_{k}(\mathbf{r})\chi_{l}(\mathbf{r})\,d\mathbf{r}}
+ \iint
        {\frac{\chi_{p}(\mathbf{r})\chi_{q}(\mathbf{r})\chi_{k}(\mathbf{r}')\chi_{l}(\mathbf{r}')}{\left| \mathbf{r}-\mathbf{r}'\right|}\,d\mathbf{r}\,d\mathbf{r}'}.
\end{align}
Hence, the iteration formula Eq.~(\ref{eq:iteration}) can be written as 
\begin{align}
    \delta\rho_{\text{damped},ij}^{\sigma} 
    = t\sum_{\sigma'kl}{K_{ij,kl}^{\sigma\sigma'}\,\delta\rho_{kl}^{\sigma'}}
    + (1-t)\delta\rho_{ij}^{\sigma}.
    \label{eq:iteration_basis}
\end{align} 
where $\{\delta\rho_{ij}^{\sigma}\}$ satisfies the constraint Eq.~(\ref{eq:sum_rule_basis}); $\{K_{ij,kl}^{\sigma\sigma'}\}$ is the undamped iteration matrix, corresponding to the operator $K$ in Eq.~(\ref{eq:iteration_operator_undamped}).

\subsection{\label{sec:limit} High-Fictitious-Temperature Limit} 

As the fictitious temperature $\theta$ increases, the unphysical spin-symmetry breaking tends to vanish in TAO-DFT \cite{Chai_2012}. In the follows, we prove that it holds for all electronic systems. 
First, we observe the asymptotic behavior of $\{\alpha_{ij}(\theta)\}$ and $\{\beta_{ij}(\theta)\}$ in the limit of high fictitious temperature. 

\begin{enumerate}
\item 
By Lagrange's mean value theorem, for any $i$ and $j$, there exists $\tilde{\varepsilon}_{ij} \in (\varepsilon_{i}, \varepsilon_{j})$ such that 
\begin{align} 
f'(\tilde{\varepsilon}_{ij}) = \frac{f(\varepsilon_{i}) - f(\varepsilon_{j})}{\varepsilon_{i} - \varepsilon_{j}}, 
\end{align} 
where $f$ denotes the Fermi-Dirac distribution function, given by Eq.~(\ref{eq:scf_f}); $f'$ denotes its derivative. 
Since
\begin{align} 
f'(\tilde{\varepsilon}_{ij}) 
= -\frac{1}{\theta} \frac{\exp[(\tilde{\varepsilon}_{ij} - \mu)/\theta]}{\{1 + \exp[(\tilde{\varepsilon}_{ij} - \mu)/\theta]\}^2} 
= -\frac{1}{\theta} f(\tilde{\varepsilon}_{ij}) [1 - f(\tilde{\varepsilon}_{ij})], 
\end{align} 
by Eq.~(\ref{eq:alpha}) and the decreasing of $f(\varepsilon)$, we obtain 
\begin{align} 
0 
\leq |\alpha_{ij}| 
\leq \left|\frac{f_i - f_j}{\varepsilon_i - \varepsilon_j}\right| 
= \frac{1}{\theta} f(\tilde{\varepsilon}_{ij})[1 - f(\tilde{\varepsilon}_{ij})] 
< \frac{f_{0}}{\theta}, 
\label{eq:alpha_asymp} 
\end{align} 
where $f_{0}$ is the maximum converged spin-restricted occupation number. Eq.~(\ref{eq:alpha_asymp}) gives
\begin{align} 
{\lim_{\theta \to \infty}{\alpha_{ij}(\theta)}} = 0. 
\label{eq:alpha_limit} 
\end{align} 

\item 
Since for any $i$ and $j$,
\begin{align} 
0 \leq&\ \frac{f_{j} \left( {1 - f_{j}} \right)}{\sum_{k}{f_{k} \left( {1 - f_{k}} \right)}} \leq 1, 
\end{align} 
by Eq.~(\ref{eq:beta}), we have 
\begin{align} 
0 
\leq \left| \beta_{ij} \right| 
=&\ \frac{1}{\theta} f_{i} \left( {1 - f_{i}} \right)\left| {\frac{f_{j} \left( {1 - f_{j}} \right)}{\sum_{k} {f_{k} \left( {1 - f_{k}} \right)}} - \delta_{ij}} \right| \notag \\ 
\leq&\ \frac{1}{\theta} f_{i} \left( {1 - f_{i}} \right)\left[ {\frac{f_{j} \left( {1 - f_{j}} \right)}{\sum_{k} {f_{k} \left( {1 - f_{k}} \right)}} + \delta_{ij}} \right] \notag \\ 
<&\ \frac{2f_{0}}{\theta}. 
\label{eq:beta_asymp} 
\end{align} 
Eq.~(\ref{eq:beta_asymp}) gives
\begin{align} 
{\lim_{\theta \to \infty}{\beta_{ij}(\theta)}} = 0. 
\label{eq:beta_limit} 
\end{align} 
\end{enumerate} 

The vanishing $\{\alpha_{ij}(\theta)\}$ and $\{\beta_{ij}(\theta)\}$ at high fictitious temperatures, according to Eq.~(\ref{eq:kernel_basis}), are the key factors 
that reduce the eigenvalues of the undamped iteration matrix $[K_{ij,kl}^{\sigma\sigma^\prime}]$. 
Since $[S_{ij}]$ in Eq.~(\ref{eq:matrix_S}) is the Gram matrix of a set of independent functions $\{\chi_i(\mathbf{r})\}$, it is positive-definite. 
Thus, for any $i$, 
\begin{align} 
1 
= {\int{|{\psi_{i}(\mathbf{r})}|^{2}d\mathbf{r}}} 
= {\sum_{jk} {c_{ij}^{*}c_{ik}S_{jk}}} 
\geq \lambda_{\text{min}}(S)\,{\sum_{j}|c_{ij}|^{2}}, 
\end{align}
where $\lambda_{\text{min}}(S) > 0$ is the minimum eigenvalue of $\left[S_{ij}\right]$. 
Then we obtain the bound of the coefficients $\{c_{ij}\}$, 
\begin{align} 
\left| c_{ij} \right| 
\leq \sqrt{\sum_{k}|c_{ik}|^{2}} 
\leq \frac{1}{\sqrt{\lambda_{\text{min}}(S)}} 
= c_{\text{bound}}, 
\end{align}
which is independent of $\theta$.
In exact functional case, $\{\xi_{pq,kl}^{\sigma\sigma'}\}$ has a bound $\xi_{\text{bound}}$ for TAO-DFT without $E_{\theta}$ since it contains no $\theta$-dependence.
By Eqs.~(\ref{eq:alpha_asymp}) and (\ref{eq:beta_asymp}), as $\theta \to \infty$, we have 
\begin{align} 
\left| K_{ij,kl}^{\sigma\sigma'} \right| 
\leq&\ {\sum_{mnpq}\left| 
    \left(\alpha_{mn}c_{mi}^{*}c_{nq}^{*}+\beta_{mn}c_{mq}^{*}c_{ni}^{*}\right)c_{mp}c_{nj}\xi_{pq,kl}^{\sigma\sigma'}
    \right|} \notag\\ 
\leq&\ \sum_{mnpq}{c_{\text{bound}}^{4} (|\alpha_{mn}| + |\beta_{mn}|) |\xi_{pq,kl}^{\sigma\sigma'}|} \notag\\
\leq&\ M_{\text{basis}}^{4}c_{\text{bound}}^{4} \frac{3f_{0}}{\theta} \xi_{\text{bound}}
\to 0, 
\end{align}
where $M_{\text{basis}}$ is the number of basis functions.
That is, 
\begin{align} 
\lim_{\theta \to \infty} K_{ij,kl}^{\sigma\sigma'} = 0. 
\end{align} 
As a result, as $\theta \to \infty$, we have 
\begin{align}
    \lim_{\theta \to \infty} \lambda = 0.
    \label{eq:asymp_lambda}
\end{align}
Eq.~(\ref{eq:asymp_lambda}) indicates that Eq.~(\ref{eq:criterion}) holds above some fictitious temperature $\theta_{\text{c}}$. 
As long as $\lambda$ becomes small enough before $E_{\theta}$ becoming significant, Eq.~(\ref{eq:criterion}) still holds above $\theta_{\text{c}}$.
Furthermore, since that $\alpha,\beta\sim f_0/\theta$ (Eq.~(\ref{eq:alpha_asymp}) and (\ref{eq:beta_asymp})) and that $E_{\theta} = A_{\text{S}}^{\theta=0} - A_{\text{S}}^{\theta} = \theta S_{\text{S}}^{\theta}$ in the exact theory (See \cite{Chai_2012}), the $E_{\theta}$ contribution in $K$ vanishes if $f_{0}S_{\text{S}}^{\theta}\to0$ as $\theta\to\infty$.
At very high $\theta$, the orbitals are populated equally and the occupation numbers equal $1/M_{\text{basis}}$; this results in
\begin{align}
    f_{0}S_{\text{S}}^{\theta}
    = \frac{\ln{M_{\text{basis}}}}{M_{\text{basis}}},
\end{align}
which is small for large $M_{\text{basis}}$. These are the reason why Eq.~(\ref{eq:criterion}) still hold for TAO-DFT with $E_{\theta}$.
According to the criterion, the spin-symmetry breaking vanishes above this critical fictitious temperature $\theta_{\text{c}}$. 

The critical fictitious temperature $\theta_{\text{c}}$ exists under the limit $M_{\text{basis}}\to\infty$ based on the following analysis of Eq.~(\ref{eq:kernel_basis}):
(i) By orthogonality of the orbitals, each $c_{ij}$ contributes to $M_{\text{basis}}^{-1/2}$ dependence;
(ii) The sum over $p,q$ contributes to $M_{\text{basis}}^{2}$ dependence;
(iii) The sum over $m,n$ does not affect any exponent of $M_{\text{basis}}$ since by Eqs.~(\ref{eq:alpha_asymp}) and (\ref{eq:beta_asymp}), both of the coefficients $\{\alpha_{mn}\}$ and $\{\beta_{mn}\}$ are bounded by $2f(\tilde{\varepsilon}_{mn})[1-f(\tilde{\varepsilon}_{mn})]/\theta$ and tend to vanish for large $m$ or $n$.
Thus, the $M_{\text{basis}}$ dependence in each term of $K$ (and thus $\lambda$) compensates with each other.
This enables us to characterize a system by its $\theta_{\text{c}}$ under a sufficiently large basis set, while different $\theta_{\text{c}}$ can be taken in the calculations with different basis sets to restore the spin-symmetry.

In short, under the assumptions 
(i) if the converged spin-restricted density is a local minimum of energy functional, it is also a global minimum, at least for a sufficiently large $\theta$ (beginning of Sec.~\ref{sec:criterion}); 
(ii) the orbitals are equally populated under high $\theta$ limit (previous paragraph), 
TAO-DFT is able to resolve the unphysical spin-symmetry breaking by a well-chosen value of $\theta$.

\section{\label{sec:calculation} Numerical Investigation} 

The analysis provided in Sec.\ \ref{sec:theory} is implemented, and examined on several molecular systems, including the dissociation of H$_{2}$, N$_{2}$, He$_{2}$, and Ne$_{2}$, as well as the twisted ethylene. 
All results are computed using TAO-LDA (i.e., TAO-DFT with the LDA xc$\theta$ energy functional) \cite{Chai_2012} with the 6-31G(d) basis set. KS-LDA (i.e., KS-DFT with the LDA xc energy functional) corresponds to 
TAO-LDA (with $\theta = 0$). The single-point calculations are performed by \textsf{Q-Chem 4.0} \cite{QC4}, whereas $\lambda$, the largest real part of eigenvalues of $K$ that contributes to spin-symmetry breaking, is obtained according 
to Sec.~\ref{sec:theory} with the converged spin-restricted data as input.

\subsection{H$_{2}$ Dissociation}

\begin{figure} 
\includegraphics[scale=1.1]{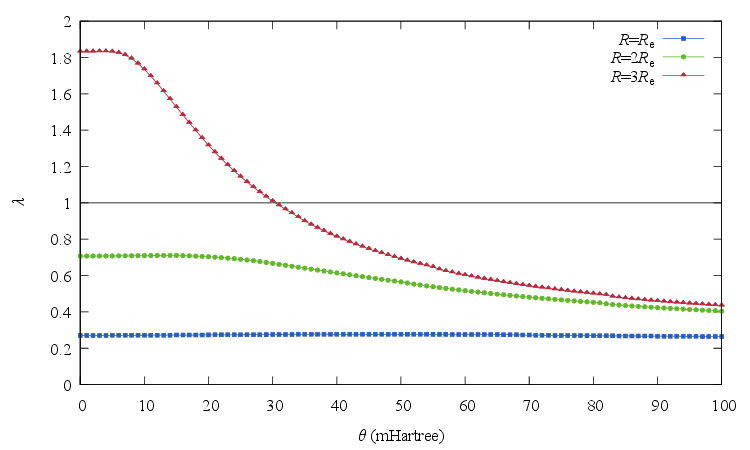} 
\caption{\label{fig:h2} 
The largest real part of eigenvalues of $K$ that contributes to spin-symmetry breaking, $\lambda$, for the ground state of stretched H$_{2}$ (with various bond lengths) 
as a function of the fictitious temperature $\theta$, calculated by TAO-LDA. The $\theta = 0$ case corresponds to KS-LDA. 
The experimental bond length $R_\text{e} = 0.741$ {\AA} \cite{Huber_Herzberg_1979,Helgaker_Jorgensen_Olsen_2000} is adopted. 
Points above the black line ($\lambda = 1$) correspond to the spin-symmetry breaking cases.}
\end{figure}

H$_{2}$ dissociation, a single-bond breaking system, is incorrectly predicted by KS-LDA. In the exact theory, due to spin symmetry, the spatial distribution of the two spin densities should be 
identical \cite{Helgaker_Jorgensen_Olsen_2000}. As shown in Fig.~\ref{fig:h2}, for the KS-LDA case ($\theta = 0$ in the figure), at the experimental bond length 
$R_\text{e} = 0.741$ {\AA} \cite{Huber_Herzberg_1979,Helgaker_Jorgensen_Olsen_2000} and $2R_\text{e}$, we have $\lambda < 1$, which means that the two spin densities are the same. However, H$_{2}$ exhibits 
spin-symmetry breaking (i.e., violation of criterion Eq.~(\ref{eq:criterion})) when the bond stretches to $3R_\text{e}$; this deviates from the exact theory. This unphysical behavior can be removed in TAO-DFT. When the 
fictitious temperature $\theta$ is above $31\,\text{mHartree}$, our theory predicts $\lambda < 1$ for the $3R_\text{e}$ case. It indicates the vanishing of spin-symmetry breaking.

\subsection{N$_{2}$ Dissociation} 

\begin{figure} 
\includegraphics[scale=1.1]{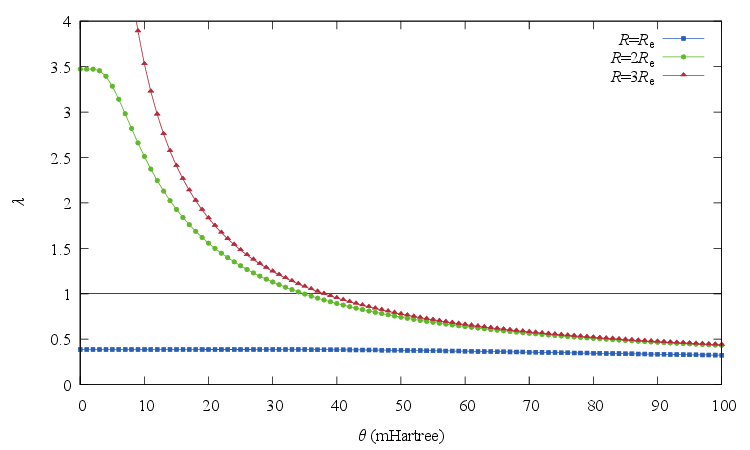} 
\caption{\label{fig:n2} 
The largest real part of eigenvalues of $K$ that contributes to spin-symmetry breaking, $\lambda$, for the ground state of stretched N$_{2}$ (with various bond lengths) 
as a function of the fictitious temperature $\theta$, calculated by TAO-LDA. The $\theta = 0$ case corresponds to KS-LDA. 
The experimental bond length $R_\text{e} = 1.098$ {\AA} \cite{Huber_Herzberg_1979,Helgaker_Jorgensen_Olsen_2000} is adopted. 
Points above the black line ($\lambda = 1$) correspond to the spin-symmetry breaking cases.}
\end{figure}

N$_{2}$ dissociation, a triple-bond breaking system, provides similar results. As shown in Fig.~\ref{fig:n2}, for the KS-LDA case ($\theta = 0$ in the figure), at the experimental bond length 
$R_\text{e} = 1.098$ {\AA} \cite{Huber_Herzberg_1979,Helgaker_Jorgensen_Olsen_2000}, we have $\lambda < 1$, which means that the two spin densities are identical. Nonetheless, N$_{2}$ exhibits spin-symmetry 
breaking (i.e., violation of criterion Eq.~(\ref{eq:criterion})) when the bond stretches to $2R_\text{e}$ and $3R_\text{e}$; this deviates from the exact theory. This unphysical behavior can be removed in TAO-DFT. When the 
fictitious temperature $\theta$ is above $38\,\text{mHartree}$, our theory predicts $\lambda < 1$ for both the $2R_\text{e}$ and $3R_\text{e}$ cases. It indicates the vanishing of spin-symmetry breaking. The dissociation 
of H$_{2}$ and N$_{2}$ justifies the theoretical high-fictitious-temperature limit in Sec.~\ref{sec:limit}.

\subsection{He$_{2}$ Dissociation} 

\begin{figure} 
\includegraphics[scale=1.1]{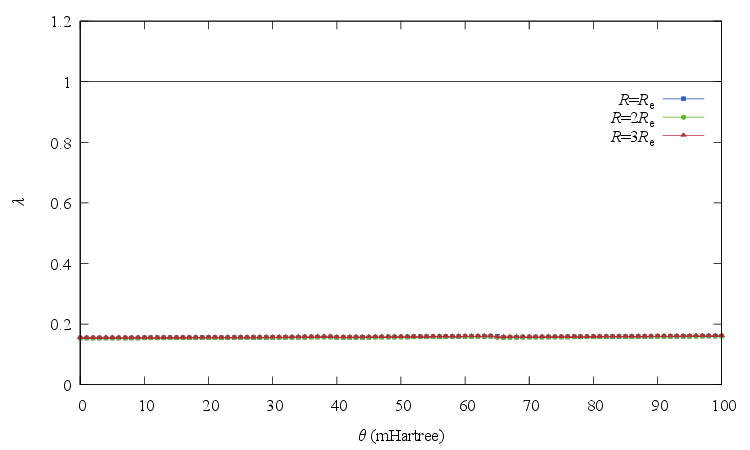} 
\caption{\label{fig:he2} 
The largest real part of eigenvalues of $K$ that contributes to spin-symmetry breaking, $\lambda$, for the ground state of stretched He$_{2}$ (with various bond lengths) 
as a function of the fictitious temperature $\theta$, calculated by TAO-LDA. The $\theta = 0$ case corresponds to KS-LDA. 
The experimental bond length $R_\text{e} = 2.967$ {\AA} \cite{Klein_Aziz_1984,Tang_Toennies_1986} is adopted. 
Points above the black line ($\lambda = 1$) correspond to the spin-symmetry breaking cases.}
\end{figure}

A stretched helium dimer (He$_{2}$) is also investigated. Since helium is a noble gas, there is no bond formation between the two helium atoms. Instead, the van der Waals (vdW) force is the reason of the attraction. 
Theoretically, the spin-symmetry breaking problem does not emerge, since each atom in the helium dimer preserves its atomic orbitals due to the lack of bond. As shown in Fig.~\ref{fig:he2}, for the KS-LDA case 
($\theta = 0$ in the figure), at the experimental bond length $R_\text{e} = 2.967$ {\AA} \cite{Klein_Aziz_1984,Tang_Toennies_1986}, we have $\lambda < 1$, 
which means that the two spin densities are equal. This result retains as the dimer distance stretches to $2R_\text{e}$ and $3R_\text{e}$. In TAO-DFT, at each dimer distance, $\lambda < 1$ always holds for any $\theta$, 
meaning the nonexistence of the spin-symmetry breaking.

\subsection{Ne$_{2}$ Dissociation} 

\begin{figure} 
\includegraphics[scale=1.1]{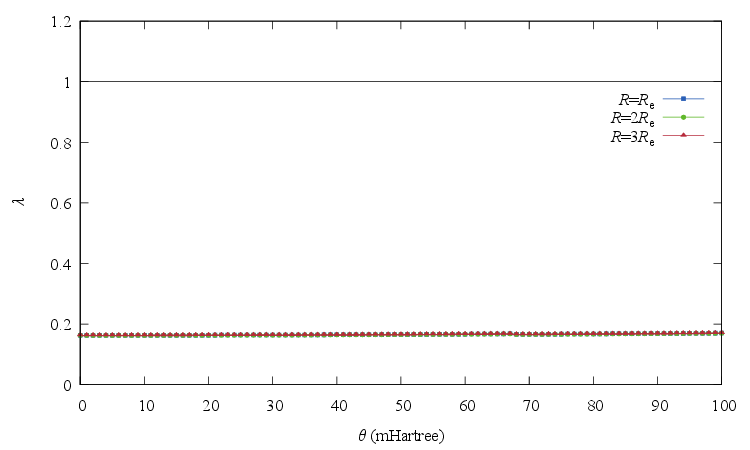} 
\caption{\label{fig:ne2} 
The largest real part of eigenvalues of $K$ that contributes to spin-symmetry breaking, $\lambda$, for the ground state of stretched Ne$_{2}$ (with various bond lengths) 
as a function of the fictitious temperature $\theta$, calculated by TAO-LDA. The $\theta = 0$ case corresponds to KS-LDA. 
The experimental bond length $R_\text{e} = 3.091$ {\AA} \cite{Aziz_Slaman_1989,Gdanitz_2001} is adopted. 
Points above the black line ($\lambda = 1$) correspond to the spin-symmetry breaking cases.}
\end{figure}

A stretched neon dimer (Ne$_{2}$) is also investigated (i.e., for another noble gas). As shown in Fig.~\ref{fig:ne2}, for the KS-LDA case ($\theta = 0$ in the figure), at the experimental bond length 
$R_\text{e} = 3.091$ {\AA} \cite{Aziz_Slaman_1989,Gdanitz_2001}, we have $\lambda < 1$, which means that the two spin densities are equal. This result retains as the dimer distance stretches to $2R_\text{e}$ and 
$3R_\text{e}$. In TAO-DFT, at each dimer distance, $\lambda < 1$ always holds for any $\theta$, meaning the nonexistence of the spin-symmetry breaking. The stretched helium dimer and neon dimer are non-MR systems, 
still justifying the theoretical high-fictitious-temperature limit in Sec.~\ref{sec:limit}.

\subsection{Twisted Ethylene} 

\begin{figure} 
\includegraphics[scale=1.1]{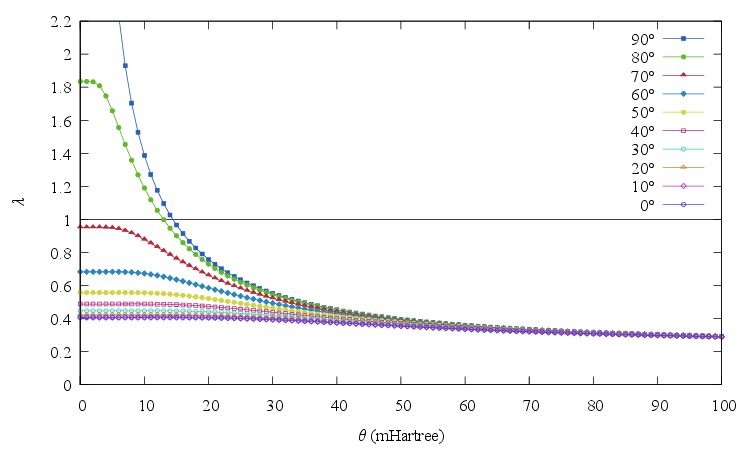} 
\caption{\label{fig:c2h4} 
The largest real part of eigenvalues of $K$ that contributes to spin-symmetry breaking, $\lambda$, for the ground state of twisted ethylene (with various HCCH torsion angles) 
as a function of the fictitious temperature $\theta$, calculated by TAO-LDA. The $\theta = 0$ case corresponds to KS-LDA. 
The experimental geometry ($R_{\text{CC}} = 1.334$ {\AA}, $R_{\text{CH}} = 1.081$ {\AA}, and 
$\angle_\text{HCH} = 117.4$$^{\circ}$) \cite{Duncan_1974,Helgaker_Jorgensen_Olsen_2000} is adopted. 
Points above the black line ($\lambda = 1$) correspond to the spin-symmetry breaking cases.}
\end{figure}

Besides the aforementioned molecular dissociation, our theory is also examined on a twisted ethylene (C$_{2}$H$_{4}$), an electronic system which has previously been shown to possess MR character \cite{Chai_2012}. 
As the HCCH torsion angle approaches 90$^{\circ}$, the $\pi$ bond between the carbon atoms breaks. As shown in Fig.~\ref{fig:c2h4}, for the KS-LDA case ($\theta = 0$ in the figure), at the experimental geometry 
($R_{\text{CC}} = 1.334$ {\AA}, $R_{\text{CH}} = 1.081$ {\AA}, and $\angle_\text{HCH} = 117.4$$^{\circ}$) \cite{Duncan_1974,Helgaker_Jorgensen_Olsen_2000}, we have $\lambda < 1$, which means the spin-up density 
is equal to the spin-down density. However, the twisted ethylene exhibits spin-symmetry breaking (i.e., violation of criterion Eq.~(\ref{eq:criterion})) at the HCCH torsion angle 80$^{\circ}$ and 90$^{\circ}$; this deviates from 
the exact theory. According to the theory in Sec.~\ref{sec:limit}, this unphysical behavior can be removed in TAO-DFT. When the fictitious temperature $\theta$ is above $15\,\text{mHartree}$, our theory predicts 
$\lambda < 1$ for the HCCH torsion angle 80$^{\circ}$ and 90$^{\circ}$. It indicates the vanishing of spin-symmetry breaking. Again, this justifies the theoretical high-fictitious-temperature limit in Sec.~\ref{sec:limit}, even for 
this non-dissociating molecular system.

\section{\label{sec:conclusions} Conclusions} 

In summary, we have proposed a theory explaining the reason why TAO-DFT can resolve the spin-symmetry breaking problem that commonly occurs in MR systems when adopting KS-DFT with the conventional xc energy functionals.
Specifically, we have characterized the spin-symmetry breaking with a dimensionless variable $\lambda$ derived from a response theory in TAO-DFT; spin-symmetry breaking occurs if $\lambda > 1$, and spin symmetry is restored if $\lambda < 1$. By the asymptotic behavior of $\lambda$, we have proved that the unphysical spin-symmetry breaking always vanishes for any system in the high-fictitious-temperature limit.
That is, for an arbitrary system, the spin symmetry can be restored in TAO-DFT by a well-chosen fictitious temperature. Besides, the theory is examined by the numerical calculations on several molecular systems, including the dissociation of H$_{2}$, 
N$_{2}$, He$_{2}$, and Ne$_{2}$, as well as the twisted ethylene. In all the cases, it has been shown that the spin symmetry can always be restored at high fictitious temperatures. 

These findings suggest the use of TAO-DFT instead of KS-DFT for MR systems. Moreover, the critical fictitious temperature that restores the spin symmetry, which exists for any system, can be chosen as the fictitious 
temperature in TAO-DFT. Once the spin-symmetry breaking problem is solved, a more accurate result in spin DFT is obtained under a similar computational cost.

\begin{acknowledgments} 

This work was supported by the Ministry of Science and Technology of Taiwan (Grant No.\ MOST110-2112-M-002-045-MY3), National Taiwan University, and the National Center for Theoretical Sciences of Taiwan. 

\end{acknowledgments}

\newpage 

\newpage 

\newpage 

\newpage 

\newpage 

\end{document}